\documentstyle[12pt]{article}

\newcommand{\e}{\; {\rm e} }
\newcommand{\sgn}{\; {\rm sgn} }
\newcommand{\tr}{\; {\rm tr} }
\newcommand{\be}{\begin{eqnarray} }
\newcommand{\ee}{\end{eqnarray} }

\begin{document}
\begin{center}
{\Large \bf Chern-Simons term  at finite density}\\[5mm]
{ \large A.N.Sissakian$\;^{a}$, \large O.Yu.Shevchenko$\;^{b,}$
\footnote{E-mail address: shevch@nusun.jinr.dubna.su}
and S.B.Solganik$\;^{b,}$
\footnote{E-mail address: solganik@thsun1.jinr.dubna.su}}\\
$\;^{a}${\it Bogolubov  Laboratory of Theoretical Physics,
Joint Institute for Nuclear Research,
Dubna, Moscow region 141980, Russia}\\
$\;^{b}${\it Laboratory of Nuclear Problems, Joint Institute for Nuclear Research,
Dubna, Moscow region 141980, Russia}
\end{center}
\vskip 5mm
\begin{abstract}
The Chern-Simons topological term  coefficient is derived
at arbitrary finite density.
As it occures that $\mu^2 = m^2$ is the
crucial point for Chern--Simons.
So when $\mu^2 < m^2$
$\mu$--influence disappears and we get the usual Chern-Simons term.
On the other hand when $\mu^2 > m^2$
the Chern-Simons term
vanishes because of non--zero density of background fermions.
In particular
for massless case parity anomaly is absent at any
finite density.
This result holds in any odd dimension as in abelian
so as in nonabelian cases.
\end{abstract}
\vskip 4mm

The great number of papers  devoted to
the Chern-Simons topological term since \cite{jackiw}
and by now have appeared.
This interest is  based on variety of
significant physical effects caused by Chern-Simons secondary
characteristic class.
These are, for example, gauge particles mass generation in quantum
field theory, applications to condense matter physics  such as
the fractional quantum Hall effect and high $T_{c}$ superconductivity,
possibility of free of metric tensor theory construction
and so on.

In a conventional zero density gauge
theory it was shown [2-4]  that the Chern-Simons term
is generated in the Eulier--Heisenberg effective action
by  quantum corrections.
The main goal of this paper is to explore the dynamical
generation of the parity anomalous Chern-Simons term
at finite density in  odd dimensional  gauge theory.

In the excellent paper by Niemi \cite{ni} it was emphasized that the charge density
at $\mu \not = 0$ becomes nontopological object, i.e contains as topological
part so as nontopological one.
The charge density at $\mu \not = 0$ (nontopological, neither parity odd
nor parity even object)\footnote{For abbreviation,  speaking about parity
invariance properties of local objects, we will
keep in mind  symmetries of the corresponding action parts.}
in $QED_{3}$ at finite density
was calculated and exploited in \cite{tolpa}.
Here we are interested in effect of finite density influence
on covariant parity odd form in action leading to the
gauge field mass generation --- Chern-Simons
topological term. Deep insight on this phenomena at small densities
was done in \cite{ni,ni1}.
The result for Chern-Simons term coefficient in $QED_{3}$  is
$\left[ \sgn(m-\mu)+\sgn(m+\mu)\right]$
(see \cite{ni1}, formulas (10.19) ).
However, to get this result it was heuristicaly supposed
that at small densities index theorem could still be used and
only odd in energy part of spectral density is responsible for
parity nonconserving effect.
Because of this in \cite{ni1} it had been stressed
that the result holds only for small $\mu$. However,
as we'll see below this result holds for any values of
chemical potential.
Thus, to obtain trustful result at any values of $\mu$ one
have to use transparent and free of any restrictions on $\mu$
procedure,
which would allow to perform calculations
with arbitrary nonabelian background gauge fields.
That is why, we will use here perturbative technique for
Chern-Simons term calculation. In addition this approach
is completely covariant.

Since  the chemical potential term $\mu\bar\psi\gamma^{0}\psi$ is
odd under charge conjugation we can expect that it would
contribute to $P$ and $CP$ nonconserving quantity ---  Chern-Simons term.
As we will see, this expectation is completely justified.

The zero density approach usually is a good quantum field  approximation
when value of chemical potential is small as compared with
characteristic energy scale  of physical processes.
Nevertheless, for investigation of topological effects
it is not the case.
As we will see below, even small density presence could lead to
the principal effects.

Introduction of a chemical potential $\mu$ in a theory   corresponds
to the presence of a nonvanishing background fermion density.
It must be emphasized that the
formal addition of a chemical potential looks like a simple gauge
transformation
with the gauge function $\mu t$. However, it doesn't only shift the time
component
of a vector potential but also gives corresponding prescription for
handling Green's function poles.
The correct introduction of a chemical potential redefines
the ground state (Fermi energy),
which leads to a new spinor propagator with the proper
$\epsilon$-prescription for poles.
So, for the free spinor propagator we have
(see, for example, \cite{shur,chod})
\begin{eqnarray}
S(p;\mu)=
\frac{\tilde{\not\! p}+m}
{(\tilde{p_{0}}+i\epsilon\sgn p_0 )^2-\vec{p}^2-m^2 },
\end{eqnarray}
where $\tilde{p}=(p_0 + \mu,\vec{p})$. It is easy to see that
in $\mu =0$ case one at once
gets the usual $\epsilon$-prescription  because of the positivity of
$p_0\sgn p_0$.

Let's first consider nonabelian 3--dimensional gauge theory.
The only graphs whose P-odd parts  contribute to the
parity anomalous Chern-Simons term are shown in Fig.1.

\unitlength=1.00mm
\special{em:linewidth 0.4pt}
\linethickness{0.4pt}
\begin{picture}(80.00,50.00)
\put(25.00,35.00){\circle{10.20}}
\put(65.00,35.00){\circle{10.00}}
\put(69.00,32.00){\line(1,0){1.00}}
\put(70.00,32.00){\line(0,-1){1.00}}
\put(70.00,31.00){\line(1,0){1.00}}
\put(71.00,31.00){\line(0,-1){1.00}}
\put(71.00,30.00){\line(1,0){1.00}}
\put(72.00,30.00){\line(0,-1){1.00}}
\put(72.00,29.00){\line(1,0){1.00}}
\put(73.00,29.00){\line(0,-1){1.00}}
\put(73.00,28.00){\line(1,0){1.00}}
\put(74.00,28.00){\line(0,-1){1.00}}
\put(61.00,32.00){\line(-1,0){1.00}}
\put(60.00,32.00){\line(0,-1){1.00}}
\put(60.00,31.00){\line(-1,0){1.00}}
\put(59.00,31.00){\line(0,-1){1.00}}
\put(59.00,30.00){\line(-1,0){1.00}}
\put(58.00,30.00){\line(0,-1){1.00}}
\put(58.00,29.00){\line(-1,0){1.00}}
\put(57.00,29.00){\line(0,-1){1.00}}
\put(57.00,28.00){\line(-1,0){1.00}}
\put(56.00,28.00){\line(0,-1){1.00}}
\put(22.00,20.00){\makebox(0,0)[lb]{(a)}}
\put(63.00,20.00){\makebox(0,0)[lb]{(b)}}
\put(2.00,10.00){\makebox(0,0)[lb]{
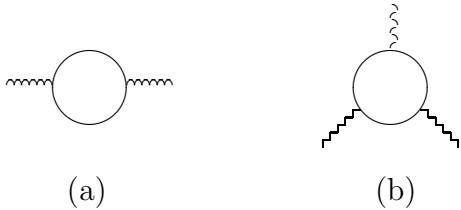
\footnotesize {Fig.1  Graphs whose P-odd parts contribute to
the Chern-Simons term in nonabelian 3D gauge theory}}}
\put(30.50,35.50){\oval(1.00,1.00)[t]}
\put(31.50,35.50){\oval(1.00,1.00)[t]}
\put(32.50,35.50){\oval(1.00,1.00)[t]}
\put(33.50,35.50){\oval(1.00,1.00)[t]}
\put(34.50,35.50){\oval(1.00,1.00)[t]}
\put(35.50,35.50){\oval(1.00,1.00)[t]}
\put(14.50,35.50){\oval(1.00,1.00)[t]}
\put(15.50,35.50){\oval(1.00,1.00)[t]}
\put(16.50,35.50){\oval(1.00,1.00)[t]}
\put(17.50,35.50){\oval(1.00,1.00)[t]}
\put(18.50,35.50){\oval(1.00,1.00)[t]}
\put(19.50,35.50){\oval(1.00,1.00)[t]}
\put(65.50,40.50){\oval(1.00,1.00)[lt]}
\put(65.50,41.50){\oval(1.00,1.00)[rt]}
\put(65.50,42.50){\oval(1.00,1.00)[lt]}
\put(65.50,43.50){\oval(1.00,1.00)[rt]}
\put(65.50,44.50){\oval(1.00,1.00)[lt]}
\put(65.50,45.50){\oval(1.00,1.00)[rt]}
\end{picture}

So, the part of effective action containing the Chern-Simons term looks as
\be
\label{eff}
I^{C.S.}_{eff} &=&
\frac{1}{2}\int_{x}  A_{\mu}(x)\int_{p}\e^{-ixp}A_{\nu}(p)
\Pi^{\mu\nu}(p)\nonumber\\ &+&
\frac{1}{3}\int_{x}  A_{\mu}(x)\int_{p,r}\e^{-ix(p+r)}
A_{\nu}(p)A_{\alpha}(r)\Pi^{\mu\nu\alpha}(p,r),
\ee
where polarization operator and vertice have a standard form
\be
\Pi^{\mu\nu}(p) &=&g^2 \int_{k} \tr \left[ \gamma^{\mu}S(p+k;\mu)
\gamma^{\nu}S(k;\mu)\right] \nonumber\\
\Pi^{\mu\nu\alpha}(p,r) &=& g^3
\int_{k}\tr \left[ \gamma^{\mu}S(p+r+k;\mu)
\gamma^{\nu}S(r+k;\mu)\gamma^{\alpha}S(k;\mu).
\right].
\ee
First consider the second order term (Fig.1, graph (a)).
It is well-known that the only object giving us the
possibility to construct $P$ and $T$ odd form in action
is Levi-Chivita tensor\footnote{In three dimensions it arises as
a trace  of three $\gamma$--matrices (Pauli matrices)}. Thus,
we will drop all terms noncontaining Levi-Chivita tensor.
Signal for the mass generation (Chern-Simons term) is
$\Pi^{\mu\nu}(p^{2}=0)\not =0$. So  we  get
\be
\Pi^{\mu\nu}=g^2 \int_{k} ( -i2m e^{\mu\nu\alpha} p_{\alpha} )
\frac{1}{(\tilde{k}^2 -m^2+i \epsilon (k_{0}+\mu)\sgn (k_{0}))^2}.
\ee
After some simple algebra one  obtains
\be
\Pi^{\mu\nu}&=&-i2mg^2e^{\mu\nu\alpha} p_{\alpha} \Biggl[\int
\frac{d^{3}k}{(2\pi)^3}
\frac{1}{(\tilde{k}^2 -m^2+i \epsilon )^2}\nonumber\\
&-& \int\frac{dk_{0}}{2\pi}\theta \left(-(k_{0}+\mu)\sgn (k_{0})\right)
\int\frac{d^{2}k}{(2\pi)^2}
\left( \frac{1}{(\tilde{k}^2 -m^2+i \epsilon )^2}-
\frac{1}{(\tilde{k}^2 -m^2-i \epsilon )^2} \right)\Biggr].
\ee
Using the well-known identity
$$\Im m \frac{1}{a\pm i\varepsilon}= \mp i\pi \delta (a)$$
and taking over integrals one immediately comes to
\be
\Pi^{\mu\nu}&=&-i\frac{m}{|m|}\frac{g^2}{4\pi}e^{\mu\nu\alpha} p_{\alpha}
\theta (m^2 -\mu^2 ) .
\ee
In the same manner handling the third order contribution (Fig.1b)
one gets
\be
\Pi^{\mu\nu\alpha} &=& -2g^3 i e^{\mu\nu\alpha}
\int\frac{d^3 k}{(2\pi)^3} \left[
\frac{m(k^2-m^2)}{ (k^2 -m^2+i\varepsilon\sgn(k_{0})(k_{0}+\mu) )^3}
\right]\nonumber\\&=&-i2mg^3  e^{\mu\nu\alpha}
\int\frac{d^3 k}{(2\pi)^3} \left[
\frac{1}{ ( k^2 -m^2+i\varepsilon\sgn (k_{0})(k_{0}+\mu) )^2 }
\right]
\ee
and  further all calculations are identical to the second order
\be
\Pi^{\mu\nu\alpha}
&=&-i\frac{m}{|m|}\frac{g^3}{4\pi}e^{\mu\nu\alpha}
\theta (m^2 -\mu^2 ).
\ee
Substituting (6), (8) in the effective action
(\ref{eff}) we get eventually
\be
I^{C.S.}_{eff} &=&\frac{m}{|m|}
\theta (m^2 -\mu^2 )\frac{g^2}{8\pi}
\int d^{3}x e^{\mu\nu\alpha} \tr\left(
A_{\mu}\partial_{\nu}A_{\alpha}-
\frac{2}{3}g A_{\mu}A_{\nu}A_{\alpha}\right).
\ee
Thus, we get Chern-Simons term with $\mu$ dependent coefficient.

Let's now consider 5--dimensional gauge theory.
Here the Levi-Chivita ten\-sor is 5--dimen\-sional $e^{\mu\nu\alpha\beta\gamma}$
and the rele\-vant graphs are shown in Fig.2.

\unitlength=1.00mm
\special{em:linewidth 0.4pt}
\linethickness{0.4pt}
\begin{picture}(120.00,50.00)
\put(25.00,35.00){\circle{10.20}}
\put(65.00,35.00){\circle{10.00}}
\put(105.00,35.00){\circle{10.00}}
\put(25.50,40.50){\oval(1.00,1.00)[lt]}
\put(25.50,41.50){\oval(1.00,1.00)[rt]}
\put(25.50,42.50){\oval(1.00,1.00)[lt]}
\put(25.50,43.50){\oval(1.00,1.00)[rt]}
\put(25.50,44.50){\oval(1.00,1.00)[lt]}
\put(25.50,45.50){\oval(1.00,1.00)[rt]}
\put(105.50,40.50){\oval(1.00,1.00)[lt]}
\put(105.50,41.50){\oval(1.00,1.00)[rt]}
\put(105.50,42.50){\oval(1.00,1.00)[lt]}
\put(105.50,43.50){\oval(1.00,1.00)[rt]}
\put(105.50,44.50){\oval(1.00,1.00)[lt]}
\put(105.50,45.50){\oval(1.00,1.00)[rt]}
\put(23.00,20.00){\makebox(0,0)[lb]{(a)}}
\put(63.00,20.00){\makebox(0,0)[lb]{(b)}}
\put(103.00,20.00){\makebox(0,0)[lb]{(c)}}
\put(3.00,10.00){\makebox(0,0)[lb]{
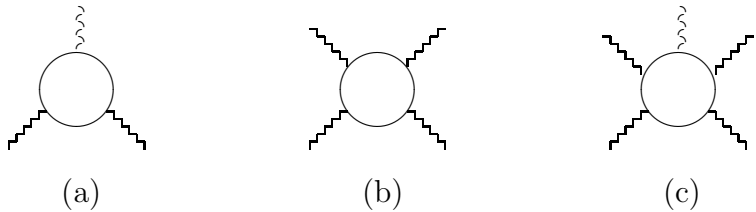
\footnotesize {Fig.2  Graphs whose P-odd parts contribute to the
Chern-Simons term in nonabelian 5D theory}}}
\put(21.00,32.00){\line(-1,0){1.00}}
\put(20.00,32.00){\line(0,-1){1.00}}
\put(20.00,31.00){\line(-1,0){1.00}}
\put(19.00,31.00){\line(0,-1){1.00}}
\put(19.00,30.00){\line(-1,0){1.00}}
\put(18.00,30.00){\line(0,-1){1.00}}
\put(18.00,29.00){\line(-1,0){1.00}}
\put(17.00,29.00){\line(0,-1){1.00}}
\put(17.00,28.00){\line(-1,0){1.00}}
\put(16.00,28.00){\line(0,-1){1.00}}
\put(29.00,32.00){\line(1,0){1.00}}
\put(30.00,32.00){\line(0,-1){1.00}}
\put(30.00,31.00){\line(1,0){1.00}}
\put(31.00,31.00){\line(0,-1){1.00}}
\put(31.00,30.00){\line(1,0){1.00}}
\put(32.00,30.00){\line(0,-1){1.00}}
\put(32.00,29.00){\line(1,0){1.00}}
\put(33.00,29.00){\line(0,-1){1.00}}
\put(33.00,28.00){\line(1,0){1.00}}
\put(34.00,28.00){\line(0,-1){1.00}}
\put(61.00,32.00){\line(-1,0){1.00}}
\put(60.00,32.00){\line(0,-1){1.00}}
\put(60.00,31.00){\line(-1,0){1.00}}
\put(59.00,31.00){\line(0,-1){1.00}}
\put(59.00,30.00){\line(-1,0){1.00}}
\put(58.00,30.00){\line(0,-1){1.00}}
\put(58.00,29.00){\line(-1,0){1.00}}
\put(57.00,29.00){\line(0,-1){1.00}}
\put(57.00,28.00){\line(-1,0){1.00}}
\put(56.00,28.00){\line(0,-1){1.00}}
\put(101.00,32.00){\line(-1,0){1.00}}
\put(100.00,32.00){\line(0,-1){1.00}}
\put(100.00,31.00){\line(-1,0){1.00}}
\put(99.00,31.00){\line(0,-1){1.00}}
\put(99.00,30.00){\line(-1,0){1.00}}
\put(98.00,30.00){\line(0,-1){1.00}}
\put(98.00,29.00){\line(-1,0){1.00}}
\put(97.00,29.00){\line(0,-1){1.00}}
\put(97.00,28.00){\line(-1,0){1.00}}
\put(96.00,28.00){\line(0,-1){1.00}}
\put(74.00,43.00){\line(-1,0){1.00}}
\put(73.00,43.00){\line(0,-1){1.00}}
\put(73.00,42.00){\line(-1,0){1.00}}
\put(72.00,42.00){\line(0,-1){1.00}}
\put(72.00,41.00){\line(-1,0){1.00}}
\put(71.00,41.00){\line(0,-1){1.00}}
\put(71.00,40.00){\line(-1,0){1.00}}
\put(70.00,40.00){\line(0,-1){1.00}}
\put(70.00,39.00){\line(-1,0){1.00}}
\put(69.00,39.00){\line(0,-1){1.00}}
\put(115.00,42.00){\line(-1,0){1.00}}
\put(114.00,42.00){\line(0,-1){1.00}}
\put(114.00,41.00){\line(-1,0){1.00}}
\put(113.00,41.00){\line(0,-1){1.00}}
\put(113.00,40.00){\line(-1,0){1.00}}
\put(112.00,40.00){\line(0,-1){1.00}}
\put(112.00,39.00){\line(-1,0){1.00}}
\put(111.00,39.00){\line(0,-1){1.00}}
\put(111.00,38.00){\line(-1,0){1.00}}
\put(110.00,38.00){\line(0,-1){1.00}}
\put(56.00,43.00){\line(1,0){1.00}}
\put(57.00,43.00){\line(0,-1){1.00}}
\put(57.00,42.00){\line(1,0){1.00}}
\put(58.00,42.00){\line(0,-1){1.00}}
\put(58.00,41.00){\line(1,0){1.00}}
\put(59.00,41.00){\line(0,-1){1.00}}
\put(59.00,40.00){\line(1,0){1.00}}
\put(60.00,40.00){\line(0,-1){1.00}}
\put(60.00,39.00){\line(1,0){1.00}}
\put(61.00,39.00){\line(0,-1){1.00}}
\put(95.00,42.00){\line(1,0){1.00}}
\put(96.00,42.00){\line(0,-1){1.00}}
\put(96.00,41.00){\line(1,0){1.00}}
\put(97.00,41.00){\line(0,-1){1.00}}
\put(97.00,40.00){\line(1,0){1.00}}
\put(98.00,40.00){\line(0,-1){1.00}}
\put(98.00,39.00){\line(1,0){1.00}}
\put(99.00,39.00){\line(0,-1){1.00}}
\put(99.00,38.00){\line(1,0){1.00}}
\put(100.00,38.00){\line(0,-1){1.00}}
\put(69.00,32.00){\line(1,0){1.00}}
\put(70.00,32.00){\line(0,-1){1.00}}
\put(70.00,31.00){\line(1,0){1.00}}
\put(71.00,31.00){\line(0,-1){1.00}}
\put(71.00,30.00){\line(1,0){1.00}}
\put(72.00,30.00){\line(0,-1){1.00}}
\put(72.00,29.00){\line(1,0){1.00}}
\put(73.00,29.00){\line(0,-1){1.00}}
\put(73.00,28.00){\line(1,0){1.00}}
\put(74.00,28.00){\line(0,-1){1.00}}
\put(109.00,32.00){\line(1,0){1.00}}
\put(110.00,32.00){\line(0,-1){1.00}}
\put(110.00,31.00){\line(1,0){1.00}}
\put(111.00,31.00){\line(0,-1){1.00}}
\put(111.00,30.00){\line(1,0){1.00}}
\put(112.00,30.00){\line(0,-1){1.00}}
\put(112.00,29.00){\line(1,0){1.00}}
\put(113.00,29.00){\line(0,-1){1.00}}
\put(113.00,28.00){\line(1,0){1.00}}
\put(114.00,28.00){\line(0,-1){1.00}}
\end{picture}

The part of effective action containing Chern-Simons term reads
\be
\label{ef}
I^{C.S.}_{eff} &=&
\frac{1}{3}\int_{x}  A_{\mu}(x)\int_{p,r}\e^{-ix(p+r)}
A_{\nu}(p)A_{\alpha}(r)\Pi^{\mu\nu\alpha}(p,r)\nonumber\\
&+&
\frac{1}{4}\int_{x}  A_{\mu}(x)\int_{p,r}\e^{-ix(p+r+s)}
A_{\nu}(p)A_{\alpha}(r)A_{\beta}(s)\Pi^{\mu\nu\alpha\beta}(p,r,s)\nonumber\\
&+&
\frac{1}{5}\int_{x}  A_{\mu}(x)\int_{p,r}\e^{-ix(p+r+s+q)}
A_{\nu}(p)A_{\alpha}(r)A_{\beta}(s)A_{\gamma}(s)\Pi^{\mu\nu\alpha\beta\gamma}
(p,r,s,q)
\ee
All calculations are similar to 3--dimensional case.
First consider third order contribution (Fig.2a)
\be
\Pi^{\mu\nu\alpha}(p,r) = g^3
\int_{k}\tr \left[ \gamma^{\mu}S(p+r+k;\mu)
\gamma^{\nu}S(r+k;\mu)\gamma^{\alpha}S(k;\mu)
\right].
\ee
Taking into account that trace of five $\gamma$-matrices in
5--dimensions is
$$
\tr\left[\gamma^{\mu}\gamma^{\nu}\gamma^{\alpha}\gamma^{\beta}\gamma^{\rho}
\right] = 4ie^{\mu\nu\alpha\beta\rho}, $$
we  extract the parity odd part of the vertice
\be
\Pi^{\mu\nu\alpha}=g^3 \int\frac{d^5 k}{(2\pi)^5}
( i4m e^{\mu\nu\alpha\beta\sigma} p_{\beta}r_{\sigma} )
\frac{1}{(\tilde{k}^2 -m^2+i \epsilon (k_{0}+\mu)\sgn (k_{0}))^3},
\ee
or in more transparent way
\be
\Pi^{\mu\nu\alpha}&=& i4mg^3 e^{\mu\nu\alpha\beta\sigma}
p_{\alpha}r_{\sigma} \Biggl[\int\frac{d^{5}k}{(2\pi)^5}
\frac{1}{(\tilde{k}^2 -m^2+i \epsilon )^3}\nonumber\\
&-& \int\frac{dk_{0}}{2\pi} \theta \left(-(k_{0}+\mu)\sgn (k_{0})\right)
\int\frac{d^{4}k}{(2\pi)^4}
\left( \frac{1}{(\tilde{k}^2 -m^2+i \epsilon )^3}-
\frac{1}{(\tilde{k}^2 -m^2-i \epsilon )^3} \right)\Biggr].
\ee
Evaluating integrals one comes to
\be
\label{p1}
\Pi^{\mu\nu\alpha}
=i\frac{m}{|m|}\frac{g^3}{16\pi^2}e^{\mu\nu\alpha\beta\sigma}
p_{\alpha}r_{\sigma} \theta (m^2 -\mu^2 ).
\ee
In the same way operating graphs (b) and (c) (Fig.2) one will obtain
\be
\Pi^{\mu\nu\alpha\beta}
=i\frac{m}{|m|}\frac{g^4}{8\pi^2}e^{\mu\nu\alpha\beta\sigma}
s_{\sigma} \theta (m^2 -\mu^2 )
\ee
and
\be
\label{p2}
\Pi^{\mu\nu\alpha\beta\gamma}
=i\frac{m}{|m|}\frac{g^5}{16\pi^2}e^{\mu\nu\alpha\beta\sigma}
\theta (m^2 -\mu^2 ).
\ee
Substituting (\ref{p1}) --- (\ref{p2}) in the effective
action (\ref{ef}) we get the
final result for  Chern-Simons in 5--dimensional theory
\be
I^{C.S.}_{eff} &=&\frac{m}{|m|}
\theta (m^2 -\mu^2 )
\frac{g^3}{48\pi^2}
\int d^{5}x e^{\mu\nu\alpha\beta\gamma} \nonumber\\
&\times&\tr\left(
A_{\mu}\partial_{\nu}A_{\alpha}\partial_{\beta}A_{\gamma}+
\frac{3}{2}g A_{\mu}A_{\nu}A_{\alpha}\partial_{\beta}A_{\gamma}+
\frac{3}{5}g^{2} A_{\mu}A_{\nu}A_{\alpha}A_{\beta}A_{\gamma}
\right).
\ee

It is remarkable that all parity odd contributions are finite
as in 3--dimensional so as in 5--dimensional cases.
Thus, all values in the effective action are renormalized in
a standard way, i.e. the renormalizations are
determined by conventional (parity even) parts of vertices.

From the above direct calculations  it is clearly seen
that the chemical potential dependent coefficient is the
same for all parity odd parts of diagrams and doesn't depend on
space dimension. So, the  influence
of finite density on Chern-Simons term generation is the same in
any odd dimension:
\begin{equation}
\label{kon}
I^{{\rm C.S}}_{eff}=\frac{m}{|m|}\theta (m^2 -\mu^2 )
 \pi W[A] ,
\end{equation}
where $W[A]$ is the Chern-Simons secondary characteristic class
in any odd dimension.
Since  only the lowest orders
of perturbative series contribute to CS term at finite density
(the same situation
is well-known at zero density), the result obtained by using
formally perturbative technique appears to be nonperturbative.
Thus, the $\mu$--dependent CS term coefficient
reveals the amazing property of universality.
Namely, it does depend on
neither dimension of the theory nor abelian or nonabelian gauge
theory is studied.

The arbitrariness of $\mu$ gives us the possibility
to see Chern--Simons coefficient behavior at any masses.
It is very interesting that  $\mu^2 = m^2$ is the
crucial point for Chern--Simons.
Indeed, it is clearly seen from (\ref{kon}) that when $\mu^2 < m^2$
$\mu$--influence disappears and we get the usual Chern-Simons term
$I^{{\rm C.S}}_{eff}= \pi W[A].$
On the other hand when $\mu^2 > m^2$
the situation is absolutely different.
One can see that here the Chern-Simons term
disappears because of non--zero density of background fermions.
We'd like to emphasize the
important massless case $m=0$ considered in \cite{ni1}.
Then even negligible density,
which always take place in any
physical processes, leads to vanishing of the parity anomaly.

In conclusion, let us stress again that we nowhere have used
any restrictions on $\mu$.
Thus we not only confirm result
in \cite{ni1} for Chern--Simons in $QED_{3}$ at small density,
but also expand it
on arbitrary $\mu$, nonabelian case and arbitrary odd dimension.

\end{document}